\documentclass{aa}

\usepackage{graphicx}
\usepackage{amsmath,amsfonts,amssymb}
\usepackage[varg]{txfonts}

\usepackage{natbib}

\begin{document}

   \title{Isotropic non-Gaussian $g_\mathrm{NL}$-like toy models that reproduce  cosmic microwave background anomalies.}

   \author{F. K. Hansen\inst{1}, T. Trombetti\inst{2},
     N. Bartolo\inst{3,4,5}, U. Natale\inst{6}, M. Liguori\inst{3,4,5},
     A. J. Banday\inst{7,8} \and K. M. G\'{o}rski\inst{9,10}}

\institute{Institute of Theoretical Astrophysics, University of Oslo, PO Box 1029 Blindern, 0315 Oslo, Norway, \email{frodekh@astro.uio.no}
\and
INAF-IRA, Bologna, Via Piero Gobetti 101, I-40129 Bologna, Italy
\and
Dipartimento di Fisica e Astronomia G. Galilei, Universit\`{a} degli Studi di Padova, via Marzolo 8, 35131 Padova, Italy
\and 
NFN, Sezione di Padova, via Marzolo 8, I-35131, Padova, Italy
\and
INAF-Osservatorio Astronomico di Padova, Vicolo dell'Osservatorio 5, I-35122 Padova, Italy
\and
Dipartimento di Fisica e Scienze della Terra, Universit\`{a} di Ferrara, Via Giuseppe Saragat 1, I-44122 Ferrara, Italy
\and
Universit\'{e} de Toulouse, UPS-OMP, IRAP, F-31028 Toulouse cedex 4, France
\and
CNRS, IRAP, 9 Av. colonel Roche, BP 44346, F-31028 Toulouse cedex 4
\and
Jet Propulsion Laboratory, California Institute of Technology, 4800
Oak Grove Drive, Pasadena, California, U.S.A.
\and
Warsaw University Observatory, Aleje Ujazdowskie 4, 00478 Warszawa, Poland
}

 \abstract
   {Based on recent observations of the cosmic microwave background (CMB), claims of statistical anomalies in the properties of the CMB fluctuations have been made. Although the statistical significance of the anomalies remains only at the $\sim2-3\sigma$ significance level, the fact that there are many different anomalies, several of which support a possible deviation from statistical isotropy, has motivated a search for models that provide a common mechanism to generate them.}{ The goal of this paper is to investigate whether these anomalies
   could originate from non-Gaussian cosmological models, and to determine what properties 
   these models should have.}{ We present a simple isotropic, non-Gaussian
   class of toy models that can reproduce six of the most extensively studied
   anomalies. We compare the presence of anomalies found in simulated maps
   generated from  the toy models and from a standard model with
   Gaussian fluctuations.} {We show that the
   following anomalies, as found in the \textit{Planck} data,
   commonly occur in the toy model maps:
   (1) large-scale hemispherical asymmetry (large-scale
   dipolar modulation), (2) small-scale hemispherical asymmetry
   (alignment of the spatial distribution of CMB power over all scales
   $\ell=[2,1500]$) , (3) a strongly non-Gaussian hot or cold spot, (4) a low
   power spectrum amplitude for $\ell<30$, including specifically (5)
   a low quadrupole and an unusual alignment between the quadrupole
   and the octopole, and (6) parity asymmetry of the lowest
   multipoles. We note that this class of toy model resembles
   models of primordial non-Gaussianity characterised by strongly
   scale-dependent $g_{NL}$-like trispectra.}{}

\authorrunning{F. K. Hansen et al.} 
\titlerunning{Non-Gaussian toy models that reproduce CMB anomalies}

\keywords{cosmology: cosmic background radiation --
  cosmology: observations -- cosmology: inflation} 

\maketitle

\section{Introduction}

Studies of the cosmic microwave background (CMB) have helped to define
the current cosmological standard model to high precision; however,  
the earliest large angular scale maps of the CMB from the COsmic Background Explorer (COBE) Differential Microwave Radiometer (DMR)
were extensively analysed to search for departures from such a model
\citep{FMG1998,PVF1998}, and then to refute them
\citep{BZG2000,Komatsu2002}. Interest in such departures,
continued with studies of the Wilkinson Microwave Anisotropy Probe (WMAP) \citep{wmap} CMB measurements,
resulting in several claims of unexpected statistical properties
(or anomalies) of the CMB fluctuations, confirmed in subsequent studies
of the \textit{Planck} data \citep{iands2013,iands2015}. While many of these anomalies
are significant only at the $2-3\sigma$ level, and could easily be the
result of statistical flukes, it is still interesting to speculate
whether they may share a common physical cosmological origin. Here, we
 investigate whether non-Gaussianity alone may be the origin of
these anomalies, including apparent deviations from statistical
isotropy and features in the power spectrum. We focus on six issues:

\begin{description}

\item [(A1)] An asymmetry of power between the two hemispheres on the sky
  was indicated by local estimates of the angular power spectrum in
  the WMAP first-year data (\citealt{asymmetry1,asymmetry2}; see also
  \citealt{varasym}).  This hemispherical asymmetry has subsequently been
  modelled by a dipolar modulation of an isotropic sky \citep{dm1,dm2},
  and detected at the $2-3\sigma$ level  for scales $\ell<60$
  in \citet{iands2015}.

\item [(A2)] While the dipolar modulation is detected only on large
  scales, the spatial distribution of power on the sky has been shown to be
  correlated over a much wider range of multipoles
  \citep{hansen2009,axelsson2013,iands2015}. By estimating the power
  spectrum in local patches of the sky for a given multipole range,
  we can create a map of the corresponding power distribution.  
  Even for an isotropic and Gaussian sky, such a map always exhibits a
  random dipole component. However, it has been shown that the directions of
  these dipole components from multipoles between $\ell=2$ to
  $\ell=1500$ are significantly more aligned in the \textit{Planck}
  data than in random Gaussian simulations. The directions of these
  dipoles are close to the direction of the best fit large-scale
   dipolar modulation in A1, but we note that A1 and A2 are two very
  distinct anomalies. A1 is present at large scales as an anomalously
  large dipolar modulation amplitude; instead, A2 is present at smaller
  scales where the amplitude of the observed dipolar modulation is
  consistent with that expected in the random Gaussian simulations, yet
  the preferred directions of the dipolar power distribution are
  aligned between multipoles. 

\item [(A3)] In \cite{vielva}, it was shown that the wavelet
  coefficients for angular scales of about $\sim10^\circ$ on the sky
  have an excess kurtosis, while the skewness is consistent with
  zero. The excess kurtosis was shown to originate from a cold spot in
  the southern Galactic hemisphere.
  However, when  the spot was masked with a disc of $5^\circ$ radius, the
  kurtosis of the map was found to be consistent with Gaussian
  simulations. The position of the cold spot on the sky lies in the
  hemisphere where the dipolar modulation in A1 is positive.
  It should also be noted that the cold spot is surrounded by a
  symmetric hot ring \citep[see][and references therein]{iands2015}.

\item [(A4)] The \textit{Planck} and WMAP power spectra of CMB temperature
  anisotropy at large scales ($\ell<30$) appear to trend
  significantly below the values predicted by the best fit
  cosmological model with a significance at the $2-3\sigma$
  level. In particular, the quadrupole is low, and a dip in the
  spectrum is observed around $\ell\sim21$. These features could be 
  statistical fluctuations on these scales where the cosmic variance
  is large.
  
\item [(A5)] The quadrupole and octopole appear to be aligned, and
  dominated by their respective high-$m$ components \citep{alignment}.

\item [(A6)] On large angular scales, the $C_\ell$ values for the even
  multipoles have been found to be consistently lower than those for
  odd multipoles.  The significance of this parity anomaly has been
  reported to be at the $2-3\sigma$ level \citep{iands2015}.

\end{description}

The correlations between some of these anomalies have been studied in
\citet{anomalycov} and shown to be largely statistically
independent.  Recent attempts in the literature to propose theoretical
explanations for anomalies have tended to focus on only one or two
examples of such behaviour, and treated them independently, with a
general emphasis on the large-scale power asymmetry.  Examples of
primordial non-Gaussianity models that have been used to explain
 the large-scale hemispherical asymmetry can be found in
\citet{schmidt2013}, \citet{byrnes2015}, \citet{byrnes2016}, \citet{shandera2016} and \citet{ashoorioon2016}. 
These models are based on earlier proposals by
\citet{Getal2005}, \citet{Erickcek2008} and \citet{DPH2008} that the properties of the
observed CMB sky could be modelled by the presence of a
long-wavelength fluctuation field that modulates otherwise isotropic
and Gaussian fluctuations. Later related studies include
\citet{Erickcek2009}, \citet{Dai}, \citet{Lyth2013}, \citet{Kanno2013}, \citet{Wang2013}, \citet{Damico2013}, \citet{McDonalda}, \citet{McDonaldb}, \citet{Liddle}, \citet{Mazu}, \citet{Mac2014}, \citet{Namjoo2013}, \citet{Namjoo2014}, \citet{Namjoo2014b}, \citet{Jaza}, \citet{Fir}, \citet{Kohri2014}, \citet{Assa}, \citet{K2015}, \citet{Agullo}, \citet{Lyth2015}, \citet{Zarei} and \citet{Tao2018}.

In particular, \citet{shandera2016} have undertaken a systematic and
general study of the power asymmetry expected in the CMB if the
primordial perturbations are non-Gaussian and exist on scales larger
than we can observe. Their analysis focuses both on local and non-local
models of primordial non-Gaussianity and the method developed is quite
general for describing deviations from statistical isotropy in a
finite sub-volume of an otherwise isotropic (but non-Gaussian) large
volume. When local non-Gaussianity is invoked, the observed scale
dependence of the power asymmetry anomaly can be recovered by the
introduction of two bispectral indices describing, on the one hand,
the scale dependence in our observable volume, and on the other hand,
a coupling to the long-wavelength fluctuation modes
\citep{schmidt2013}. In \citet{byrnes2016}, previous calculations
restricted to one- or two-source scenarios have been extended. They compute the response of the two-point function to a
long-wavelength perturbation in models characterised by a near-local
bispectrum. However, in all of these works only the effects of the
second-order terms ($f_\mathrm{NL}$) in the primordial non-Gaussianity
have been studied in detail, and the main focus has been on the
large-scale power asymmetry. Only recently, in \citet{shandera2018}, was it
 shown that large-scale power asymmetry may arise in models with
local trispectra with strong scale dependent $\tau_\mathrm{NL}$
amplitudes. However, in this case it is not possible to reproduce
all the observed CMB anomalies.  Typically, a $\tau_{\rm NL}$
trispectrum arises from the modulation of the primordial curvature
perturbation by a second uncorrelated field  
\citep[see e.g.][]{byrnes2006,gauss2013}.  As we show, this in general fails
to achieve the enhancement of and correlations to the linear
Gaussian field that are necessary ingredients to reproduce
anomalies other than the large-scale power symmetry. However, 
these features are included in our toy model.

Alternative inflationary models have  been proposed to explain CMB
anomalies such as the lack of power on large angular scales. In this
case, the models rely on deviations from the usual slow-roll phase in
a period immediately before the observable 60 e-folds. The
anomalies on the largest scales could provide hints about the
conditions that led to the inflationary dynamics (in the observable
window) given that they appear on the largest scales that will ever be
observable \citep[see
e.g.][]{infl2015,Contaldi,Liu,Gruppuso2016,Gruppuso2018}.

However, the majority of the inflationary models proposed to date to
explain the CMB anomalies have encountered difficulties
\citep{infl2015,byrnes2016,Contreras2018}.  Therefore, in this paper,
we prefer to consider that the anomalous features have a {common}
cosmological origin, and look for toy models that can naturally
reproduce all of the above anomalies.  In particular, inspired
by the additional (non-linear) terms in the primordial gravitational
potential that appear in models of inflation, we search for isotropic
but non-Gaussian models, where the non-Gaussianity is the origin of
the apparent deviations from statistical isotropy seen in the
data.  We note that the focus of this work is not to find physical
models that fit the data, but to determine phenomenologically those
properties that a physical model should exhibit.

\section{Phenomenological models}

Inflationary models may have second-order ($f_\mathrm{NL}$-like) and
third-order ($g_\mathrm{NL}$-like) terms in the primordial
gravitational potential. In the local version, these can be written
as~\citep{Gangui,Verde,Wang,Komatsu,Okamoto}
\begin{equation}
\label{eq:gnl}
\Phi(\vec{x}) = \Phi_G(\vec{x}) + f_\mathrm{NL}(\Phi^2_G(\vec{x})-\langle\Phi^2_G(\vec{x})\rangle) + g_\mathrm{NL}\Phi^3_G(\vec{x})\, , 
\end{equation}
where $\Phi_G(\vec{x})$ is the linear Gaussian part of the primordial
gravitational potential.  Clearly, models with a second-order
$f_\mathrm{NL}$ term would result in excess skewness, and not (at
lowest order) the excess kurtosis  seen in the cold spot. In order to
reproduce the latter, we will therefore focus on $g_\mathrm{NL}$-like
models. The value of the local (scale-independent) $g_\mathrm{NL}$
term has already been constrained (at the $68 \%$ confidence level) to
be $g_\mathrm{NL}=(-9.0 \pm 7.7) \times 10^4$ \citep{gauss2015}. Here,
we  instead consider $g_\mathrm{NL}$-like models with a strong
scale dependence, for which there are no current observational
constraints.  However, an indication of the level of scale-dependent
$g_\mathrm{NL}$ in the data may be found through the diagonal of the
trispectrum. We  compare the kurtosis of our models at different
scales with current observational constraints.

To motivate the construction of our toy model, we begin by considering
two related anomalies: the dipolar modulation of power at large scales
(A1) and the correlations between randomly oriented power dipoles over
a large number of angular scales (A2).  We consider the modulation of an
isotropic Gaussian CMB map,
\begin{equation}
\label{eq:dipmod}
T(\theta,\phi)=T_\mathrm{G}(\theta,\phi)(1+\beta T_\mathrm{MOD}(\theta,\phi)),
\end{equation}
where $T_\mathrm{G}(\theta,\phi)$ is an isotropic Gaussian CMB
temperature realisation, $\beta$ is the modulation amplitude, and
$T_\mathrm{MOD}(\theta,\phi)$ is the modulation field. If the
modulation field were a pure dipole, as considered in \citet{dm1}, we
would only reproduce anomaly A1. However, if we consider a modulation
field that corresponds to the original isotropic CMB map filtered such
that only the largest scales, $\ell<30$, remain (hereafter
$T_\mathrm{F}(\theta,\phi)$), then the CMB sky will have the following
features:
\begin{enumerate}
\item All scales will be correlated with the largest scales; in
  particular, the random dipolar distribution of power on the sky for
  the larger scales will be imprinted on the smaller scales giving
  rise to anomaly A2.
\item The random dipolar distribution of power on the sky for the
  larger angular scales will be enhanced, thereby mimicking a dipolar
  modulation of these scales and giving rise to anomaly A1. This
  effect is only achieved if the modulation field amplifies both the
  positive and negative fluctuations. This requires
  $T_\mathrm{MOD}(\theta,\phi)$ to be related to the absolute value of
  the filtered original map, most simply achieved by setting the
  modulation field equal to $T_\mathrm{F}^2(\theta,\phi)$.
\item A model with such a modulation field will also amplify the
  hottest and coldest spots on the map. These hot and cold spots will
  correspond to the points on the map where the non-Gaussianity
  introduced by the modulation is most easily measured. As the
  non-Gaussian term corresponds to the third power of a Gaussian
  field, it will give rise to excess kurtosis in these spots, thus
  reproducing anomaly A3. We note that no skewness can arise from a third-order term.
\end{enumerate}

\begin{figure*}[htbp]
\centering
   \hspace*{-2cm}\includegraphics[scale=0.5]{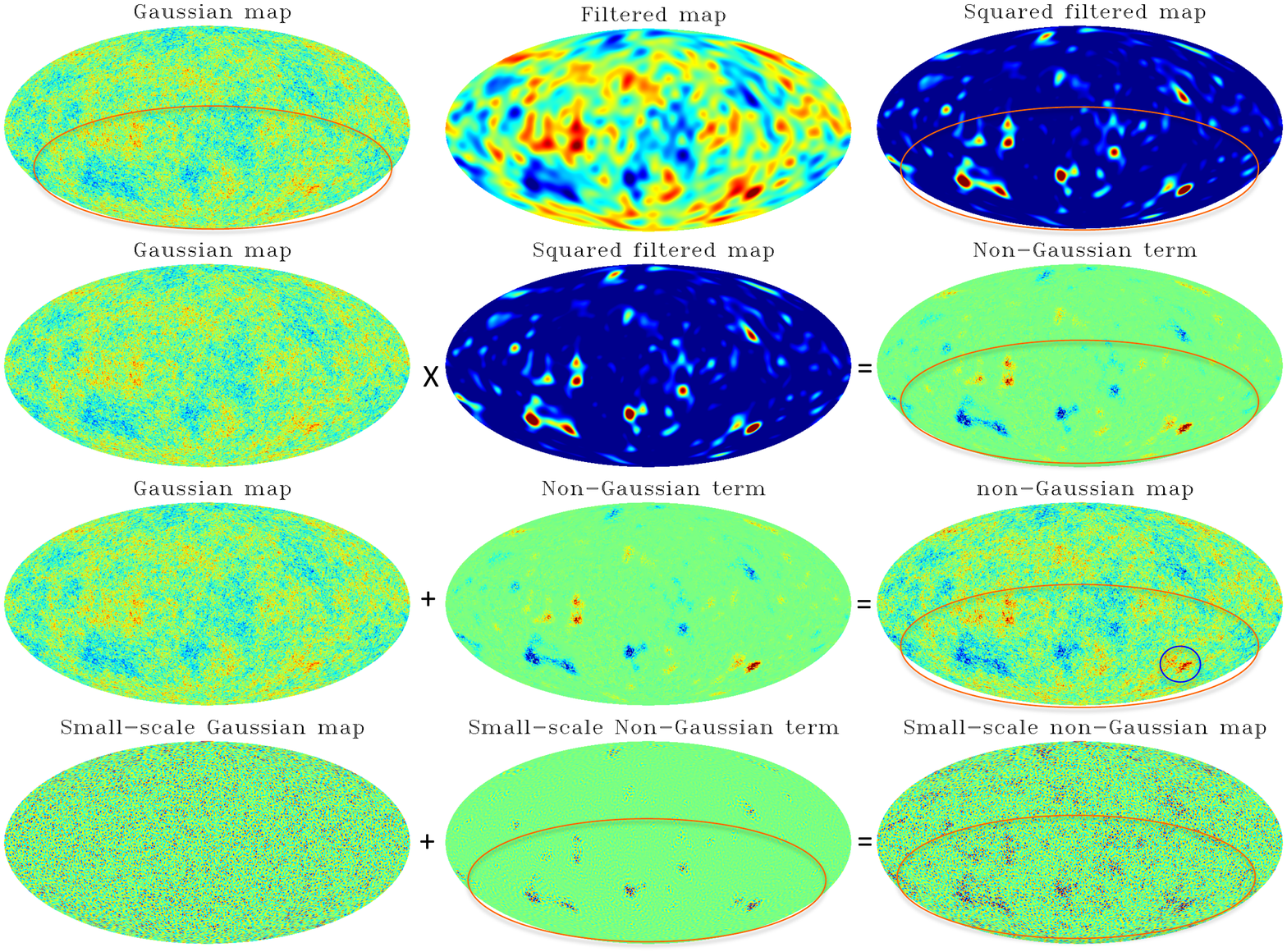}
   \caption{{\bf First row:} Gaussian CMB realisation (left), the
     filtered version (middle), and the square of the filtered map
     (right). The orange ellipse denotes the random dipole direction
     observed on large angular scales. For the squared map,
     dark blue corresponds to zero temperature, while for all other
     maps dark red and dark blue correspond to the largest negative
     and positive fluctuations in the map.  {\bf Second row:} Product of the Gaussian map (left) and the squared filtered map
     (middle)  generates the non-Gaussian contribution (right) that
     enhances the dipole in the power distribution. {\bf Third row:}
     Combination of the original Gaussian map (left) and the
     non-Gaussian term (middle) scaled by the factor
     $\beta_\mathrm{dimensionless}=1.77\times10^8$ yielding a
     non-Gaussian map (right) with enhanced dipole modulation and a
     hot spot with excess kurtosis shown in the blue circle. {\bf Last
       row:} As in the third row, but for scales $\ell=100-200$ only.  The value of $\beta$ is exaggerated here in order to make
     the effect visible by eye.}
     \label{fig:illustration}
\end{figure*}

This mechanism is illustrated in Fig~\ref{fig:illustration}. The upper row
shows a Gaussian CMB temperature realisation
$T_\mathrm{G}(\theta,\phi)$; the corresponding low-pass
filtered map $T_\mathrm{F}(\theta,\phi)$; and the square of
this filtered map,  the modulation field.  The second
row shows what happens when the modulation field is multiplied by the
Gaussian field - strong large-scale fluctuations are enhanced and the
remaining fluctuations suppressed.

We see by eye that this Gaussian realisation contains more large-scale
fluctuations in the southern hemisphere than in the northern
hemisphere (indicated by the orange oval). Such a random dipolar
distribution of power is common to all Gaussian realisations. Since the direction of such a dipolar distribution is random,
no evidence of this behaviour is seen when the mean is taken over many
simulations. The large-scale fluctuations in the southern hemisphere
are then enhanced, as shown in the second row of the figure. In the
third row, the non-Gaussian term obtained in the second row is added
to the Gaussian map, thereby enhancing the large-scale fluctuations in
the southern hemisphere and giving rise to a dipolar modulation on large
scales. We also note  the non-Gaussian hot-spot created by the
non-Gaussian term, as highlighted by the blue circle.  For this specific
realisation, the non-Gaussian feature is a hot spot since the largest
fluctuation on the sky is positive. For realisations where the
largest fluctuation is negative, a non-Gaussian cold spot will
arise. However, since the hot spot is necessarily created on top of a
hot fluctuation, a corresponding cold spot would be created on top of
a cold fluctuation and the observed feature of a hot ring
surrounding the cold spot (see anomaly A3 above) will not be created
in this simplified model. Below (Eq. \ref{eq:model}) we describe a more sophisticated model
that can reproduce all anomalies A1 to A6.

Finally, in the lowest row of Fig.~\ref{fig:illustration}, we present
maps filtered to contain the angular scales for $\ell=100-200$ only,
before and after adding the non-Gaussian term. The large-scale
structure in the southern hemisphere in the Gaussian map has been
imprinted on these smaller scales (as seen in the middle plot). Adding
this small-scale structure tilts the random dipolar distribution of
power on the sky for these scales towards the south, as in anomaly
A2. This happens for all scales.

We thus propose an initial toy model, written as
\[
T(\theta,\phi)=T_\mathrm{G}(\theta,\phi) + \beta T_\mathrm{G}(\theta,\phi)  T_\mathrm{F}^2(\theta,\phi) 
\]
to reproduce anomalies A1, A2, and A3. 
This model would leave a strong imprint of anomalies on all scales. In
order to avoid anomaly A2 becoming too pronounced, and to obtain a map
that is consistent with the observed CMB sky, the final term must
itself be filtered. We therefore modify the above model as
\begin{eqnarray}
\label{eq:model}
T(\theta,\phi)&=&T_\mathrm{G}(\theta,\phi) + \beta \left[T_\mathrm{G}(\theta,\phi)  T_\mathrm{F}^2(\theta,\phi)\right]^\mathrm{Filtered},\\
&=&T_\mathrm{G}(\theta,\phi) + \beta \sum_{\ell m} g_l Y_{\ell
  m}(\theta,\phi) \int d\Omega' Y_{\ell m}^* (\theta',\phi') \times\\
&&T_\mathrm{G}(\theta',\phi')  T_\mathrm{F}^2(\theta',\phi'), \nonumber
\end{eqnarray}
where
\begin{equation}
\label{eq:tf}
T_\mathrm{F}(\theta',\phi')=\sum_{\ell m} w_l Y_{\ell m}(\theta,\phi) \int d\Omega' Y_{\ell m}^* (\theta',\phi') T_\mathrm{G}(\theta',\phi')
\end{equation}
The filters $w_\ell$ and $g_\ell$, and the amplitude $\beta$ are then
adjusted to test whether the anomalies can be reproduced.  In addition
to this model, we also tested a variant with similar behaviour,
\begin{eqnarray}
\label{eq:altmodel}
T(\theta,\phi)&=&T_\mathrm{G}(\theta,\phi) + \beta \left[T_\mathrm{G}(\theta,\phi)  \left\{T_\mathrm{G}^2(\theta,\phi)\right\}_F\right]^\mathrm{Filtered}
\end{eqnarray}
The difference to the  original model should be noted:  the Gaussian field is
squared before the filter $w_\ell$ is applied. In this paper results
will always be based on the model specified by Eq.~\eqref{eq:model},  
unless we explicitly refer to the alternative
Eq.~\eqref{eq:altmodel}.

\begin{figure*}[htbp]
\centering
   \hspace*{-0.8cm}\includegraphics[width=0.55\linewidth]{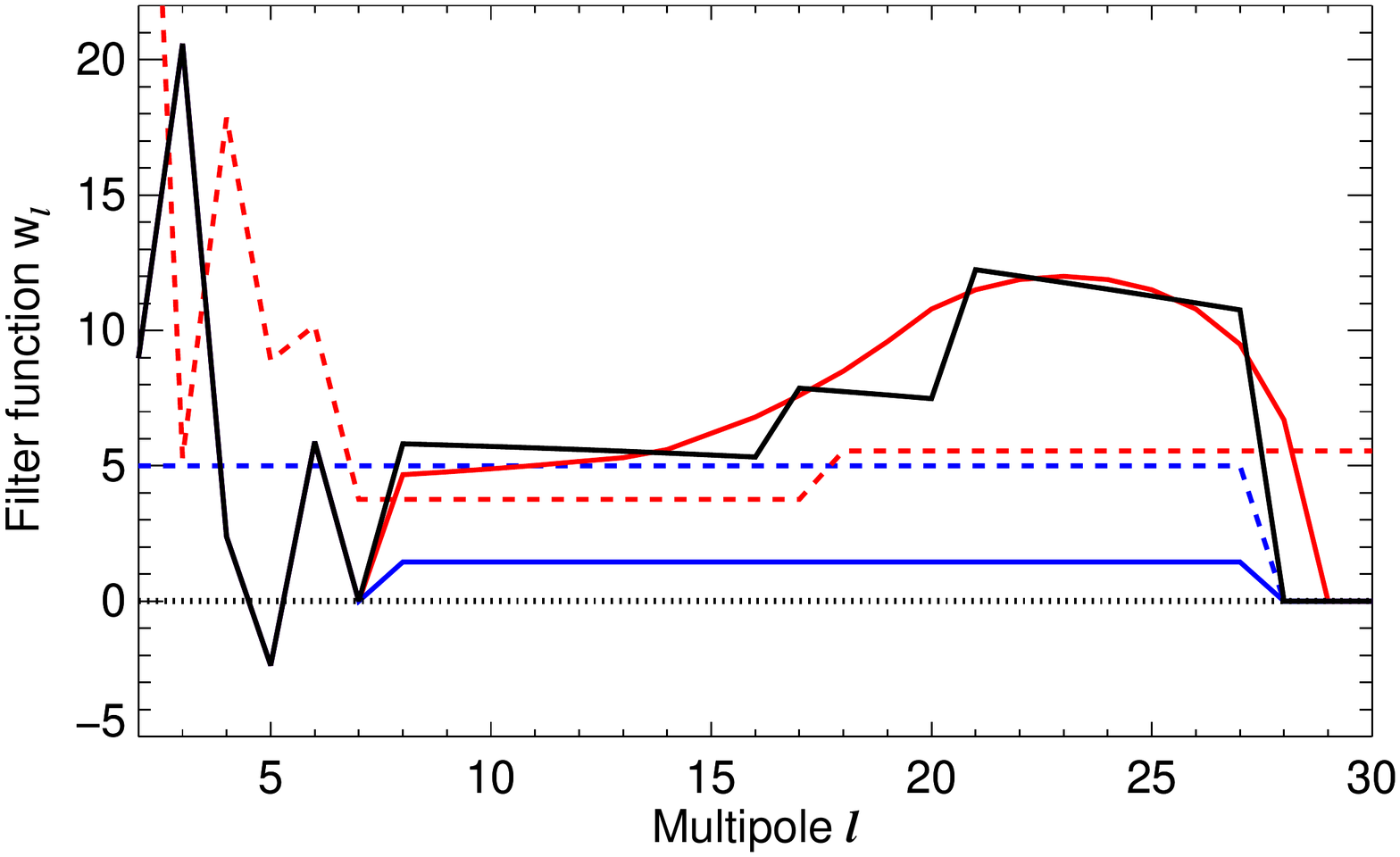}\hspace*{-1cm}\includegraphics[width=0.55\linewidth]{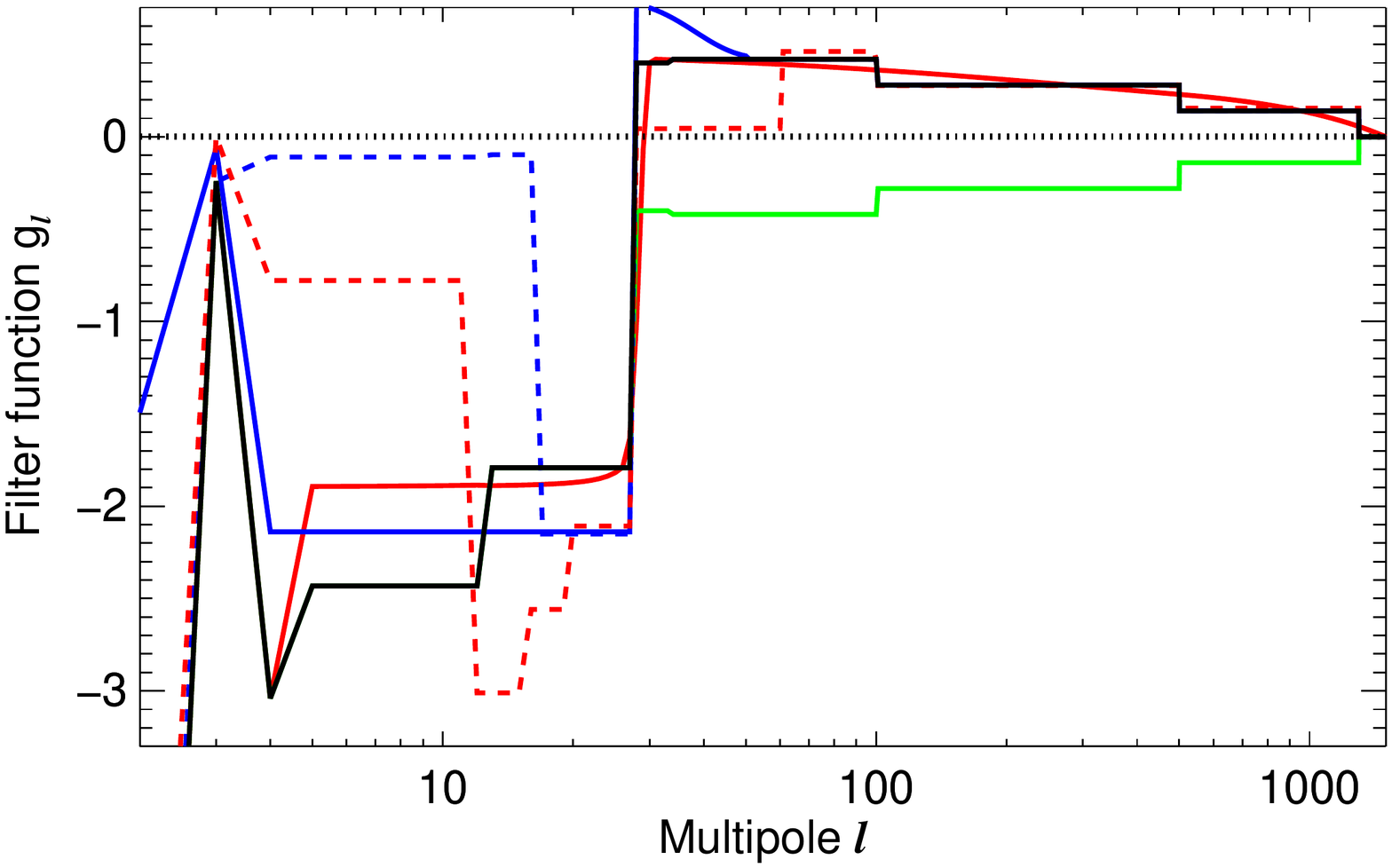}
   \caption{Filter functions $w_\ell$ (left) and $g_\ell$ (right)
     used in our toy model. For $g_\ell$, all filters except the blue
     filter have $g_\ell=-8$ for $\ell=2$, which is not shown in order
     to make the filters more visible for all other multipoles. The
     black lines show the filters that form the basis for the
     majority of results presented in the paper. The coloured filters
     are equal to the black filter for multipoles where the coloured
     filters are not visible. Red line: a smoother version of the main
     filter; green line: similar to the black filter but slightly
     simpler with no change of sign in $g_\ell$; blue dashed line:
     highly simplified $w_\ell$; red dashed line: the alternative
     model which filters the squared Gaussian map; blue line: filter
     used for the model with Gaussian white noise maps.}
     \label{fig:filters}
\end{figure*}

Figure~\ref{fig:filters} presents the filters $w_\ell$ and $g_\ell$,
shown as solid black lines, used for the majority of results in this
paper.  They correspond to one representative example of a
huge variety of filters that, to different extents, can reproduce
properties of the anomalies. Some other filters, commented on below,
are also shown. The black filters are adjusted to the form shown in
Fig.~\ref{fig:filters} in order to reproduce the shape of the power
spectrum on large scales, specifically to reproduce anomalies A4 and
A6, and to ensure that anomaly A2 is present on smaller scales.

The shape defined for the black filters in Fig.~\ref{fig:filters}
can be understood as follows. The oscillations in the lowest
multipoles of $w_\ell$ give rise to the observed parity asymmetry, and
were adjusted to reproduce the observed power spectrum oscillations
for $\ell<10$. We did not attempt to reproduce the parity asymmetry at
higher multipoles here, but given the strong correlations between
$w_\ell$ and the shape of the power spectrum, a model with wiggles in
this filter could give rise to odd-even features in the spectrum also
at other multipoles in the same way as we have shown for $\ell<10$. The filter $w_\ell$ then rises incrementally to a plateau around $\ell=21$, after
which it drops to zero at $\ell=28$. This allows the model to
reproduce the large trough at $\ell\sim21$. For $\ell>27$ the observed
power spectrum no longer lies systematically below the model spectrum,
thus $w_\ell$ can be zero for higher multipoles.  We note that the
purpose of the features of $w_\ell$ shown by the black line is to
reproduce the particular features in the power spectrum. A completely
flat $w_\ell$ filter which is zero for $\ell>27$ (as shown by the blue
dashed line) can still give rise to all anomalies except parity asymmetry (A6)
and quadrupole-octopole alignment (A5) for which a higher value of
$w_\ell$ at $\ell = 2$ and $\ell = 3$ is necessary.

The black filter $g_\ell$ in Fig.~\ref{fig:filters} is negative up to
$\ell=27$ where it suddenly becomes positive and then gradually
decreases towards zero at high $\ell$. For large scales, $g_\ell$ is
negative in order to subtract power for $\ell<27$ making the power
spectrum low for this multipole range, a positive $g_\ell$ here
instead would yield a large-scale power spectrum with high amplitude.
Then, in order to reproduce anomaly A2 on small scales, $g_\ell$ needs
to be non-zero up to $\ell\sim1500$ (but can be positive, negative, or
oscillate between the two). The filter $g_\ell$ needs to decrease
slowly to zero towards $\ell\sim1500$ in order to limit A2. The strong
negative value for $\ell=2$ in $g_\ell$ is in order to ensure a small
quadrupole, but will also strongly influence the original quadrupole
generating correlations with the octopole as in anomaly A5.

The black filters in Fig.~\ref{fig:filters} are constructed from a
combination of step functions in order to obtain the general
properties described above. A physical model would more naturally have
a smoother scale dependence, but the purpose of this paper is to
describe a toy model that presents the general features of scale
dependences that can give rise to the CMB anomalies. We do not attempt
to derive a model that can be fitted to the data here, since the
number of degrees of freedom is too large, and a theoretical model
would be needed that naturally gives rise to this scale dependence for
a minimum number of additional parameters.  Such models will be
explored in future work \citep{tutti}.

Figure~\ref{fig:filters}  presents some additional representative
examples of filters that can reproduce most or all of the anomalies.
The red lines show a smoother version of the black filter giving very
similar results. The filter shown in green differs from the black
filter in that it is negative for all multipoles, again reproducing
all anomalies. The blue dashed filter is much simpler;  since $w_\ell$
is flat, the odd-even oscillations in the power spectrum for low
multipoles are not reproduced (thus anomaly A6 disappears) and the
trough at $\ell=21$ is not visible. Even with this simple filter,
several anomalies are present. The red dashed lines represent the
filters that are used for the alternative model in
Eq.~\eqref{eq:altmodel}. The blue filter is used for Gaussian maps
replaced by white noise maps as described below. Due to limitations in
available CPU hours, only the maps based on the black filters were studied in detail using 3000 simulations, with 1000 maps
simulated for the other filters.

\begin{figure}[htbp]\vspace*{1cm}
\centering
   \hspace*{-9cm}
%\vspace*{-3cm}
{\includegraphics{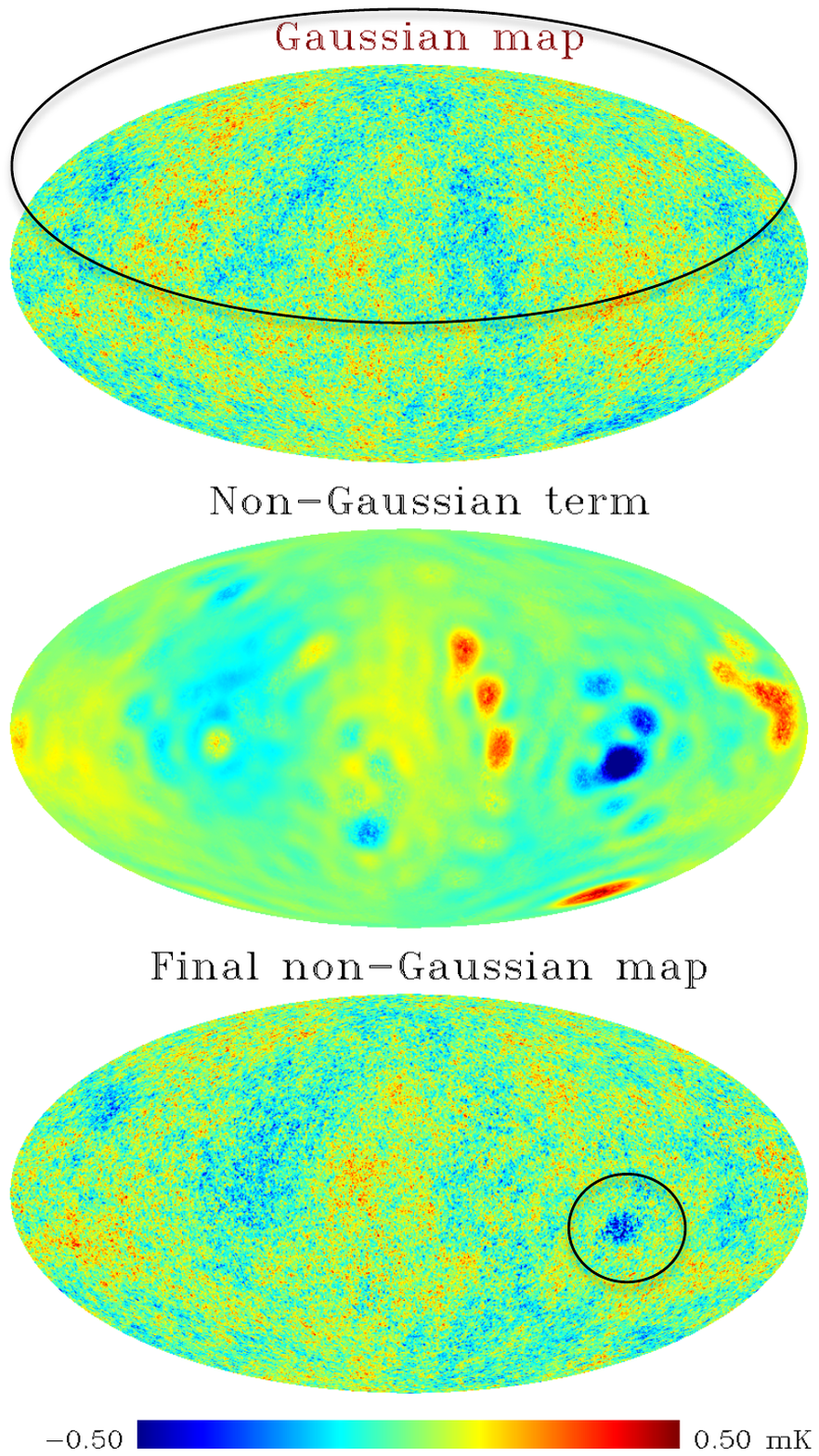}}
   \caption{{\bf Top}: Original simulated Gaussian map. The circle
     indicates the hemisphere with the most large-scale power. {\bf Middle}:
     Additional term used in our non-Gaussian model scaled by
     $\beta_\mathrm{dimensionless}=4.4\times10^6$, as used for the simulated
     model. {\bf Bottom}: Non-Gaussian map created by the
     addition of the second map to the Gaussian map. The circle highlights
     a cold spot surrounded by a hot ring.}
     \label{fig:maps}
\end{figure}

\begin{figure}[htbp]
\centering
   \includegraphics[width=0.7\linewidth]{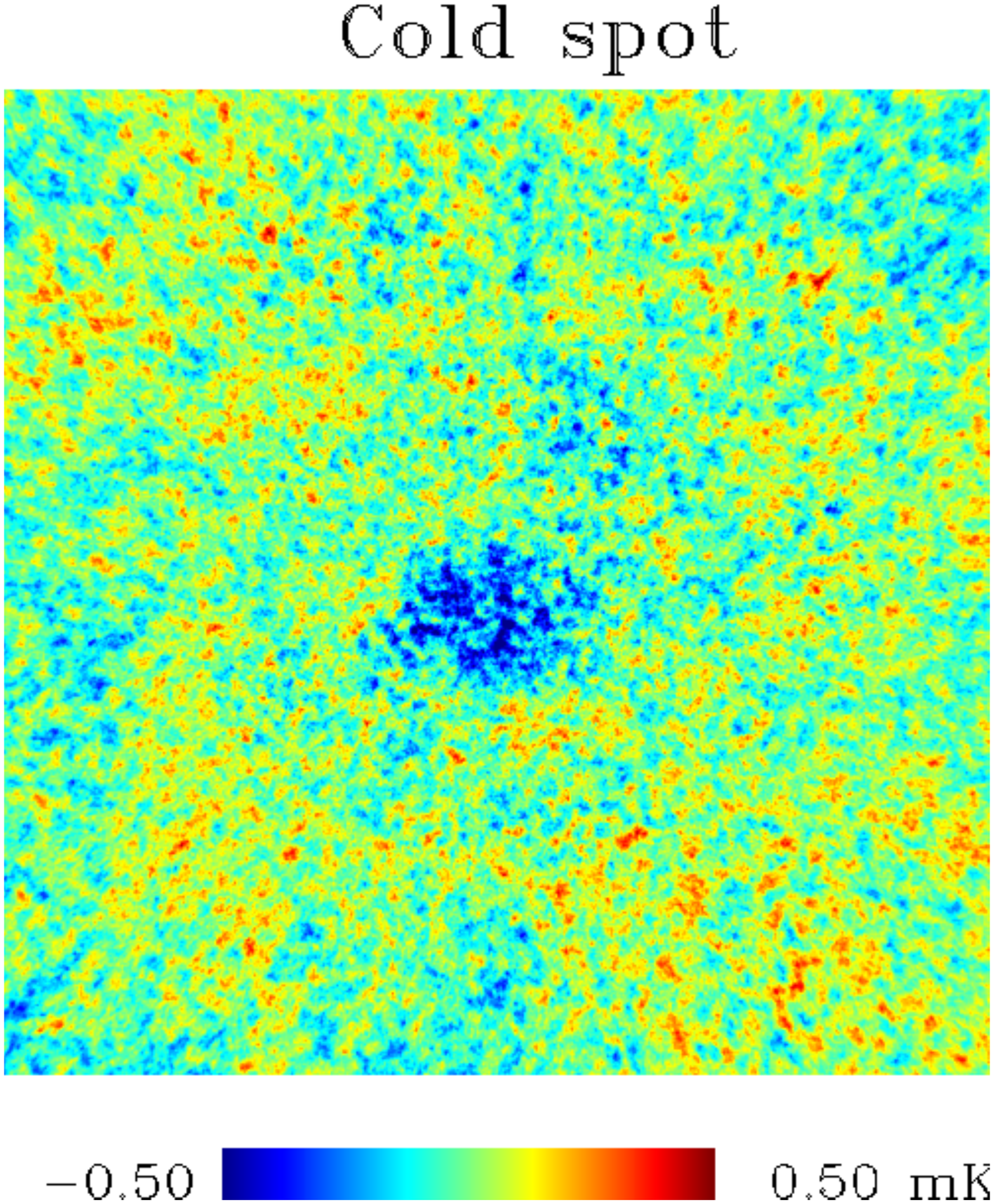}
   \caption{Cold spot with the hot ring from the simulation shown in Fig.~\ref{fig:maps}}
     \label{fig:coldspot}
\end{figure}

In Fig.~\ref{fig:maps} we show the non-Gaussian term for a simulated
map generated using the black filters in Fig.~\ref{fig:filters}. The
figure shows one of the maps from our simulation pipeline described
below where a dimensionless\footnote{Dimensionless $\beta$ refers to
  the amplitude determined when the maps are made dimensionless after
  dividing by $2.73$K in Eq.~\eqref{eq:model}.}  amplitude
$\beta_\mathrm{dimensionless}=4.4\times10^6$ is used. For this
realisation, the northern hemisphere of the Gaussian map has more
large-scale power, which is then enhanced in the non-Gaussian map. The
negative (or possibly oscillating) $g_\ell$ for larger scales makes
the non-Gaussian term more complicated and less intuitive than the
simplified illustration in Fig.~\ref{fig:illustration}. In this
case, fluctuations on some scales are enhanced and others are
suppressed. In particular, strong cold fluctuations can now appear
superposed on larger hot fluctuations and vice versa. In this way, a
cold spot can be found with a hot surrounding ring as observed in the
\textit{Planck} data (see anomaly A3 above). This is clearly seen in
Fig.~\ref{fig:coldspot} which shows a zoomed-in image of the cold spot from the
simulated map in the lower panel of Fig.~\ref{fig:maps}.

Comparing the toy model in Eq.~\eqref{eq:model} to the theoretical
$g_\mathrm{NL}$-model in Eq.~\eqref{eq:gnl}, some similarities are
apparent, but clearly a scale-dependent $g_\mathrm{NL}$ model is
required where the scale dependence defines the shape of the filters
$w_\ell$ and $g_\ell$. Unlike physical $g_\mathrm{NL}$
models, we note that a full CMB map with radiative transfer is included in all
three fields in the non-Gaussian term in Eq.~\eqref{eq:model}. As is  discussed later, the spectrum used for the Gaussian map is actually
unimportant.  The anomalies can be adequately reproduced using either
a pure Sachs-Wolfe or a white noise spectrum to generate the
Gaussian map used as an input for creating the non-Gaussian term.

The term $T_\mathrm{F}^2(\theta,\phi)$ has a dipole
($\ell=1$) component by construction, even if
$T_\mathrm{F}(\theta,\phi)$ has a zero dipole. This directly induces a
modulation in the final map, as in Eq.~\eqref{eq:dipmod}, which is
much larger than observed in the data. A physical model must therefore
have an additional filter that effectively reduces this dipolar term
(or, by coincidence, this dipole is small in the actual Universe). In
the test simulations used here, this dipole is set to zero by
hand, except for the alternative model given by
Eq.~\eqref{eq:altmodel} where it is resolved if $w_\ell$ is
low or zero for $\ell=1$. We further note that in order for the final
map to have a small-scale power spectrum that matches the data, the
total amplitude of the original map must be adjusted slightly. This
corresponds to a cosmological model with a slightly lower amplitude of
primordial fluctuations.

\section{Simulations and comparison with data}

In order to compare the probability of finding the observed anomalies
in toy model simulations to that in Gaussian simulations, we  used
a set of 3000 simulated Gaussian \textit{Planck} maps \citep{sims2015} which
were propagated through the {\tt SMICA} foreground cleaning pipeline in
order to compare them with the data cleaned with {\tt SMICA} foreground
subtraction method \citep{compsep2015}. The simulations were divided
into three sets of 1000 simulations. Set 1 was used to calibrate the
probabilities used to find $p$-values for the anomalies, set 2 was
used to create the non-Gaussian simulations, and set 3 was used to
compare Gaussian with non-Gaussian simulations.

We use anomaly A1 as an example of how these three sets were used:
\begin{enumerate}
\item The dipole modulation amplitudes (corresponding to $\beta$ in
  Eq.~\ref{eq:dipmod} with a dipole as the modulation field) were estimated for all three sets of
  simulations for a given maximum multipole $\ell_\mathrm{max}$.
\item The dipole modulation amplitude of one selected simulation in
  set 3 (Gaussian) was compared to all 1000 Gaussian simulations in
  set 1 in order to find the fraction of maps in set 1 with a larger
  amplitude than the selected simulation from set 3. This fraction is
  the $p$-value for this given set 3 simulation.
\item This procedure was repeated for all 1000 simulations in set 3 to
  determine 1000 $p$-values for a given $\ell_\mathrm{max}$.
\item Points 2 and 3 were  repeated using set 2 in  place of set 3:
  in this way we compared the dipole modulation amplitudes of all 1000
  non-Gaussian simulations in set 2 to set 1. 
\end{enumerate}

\begin{figure}[htbp]
\centering
   \hspace*{-0.6cm}\includegraphics[width=1.2\linewidth]{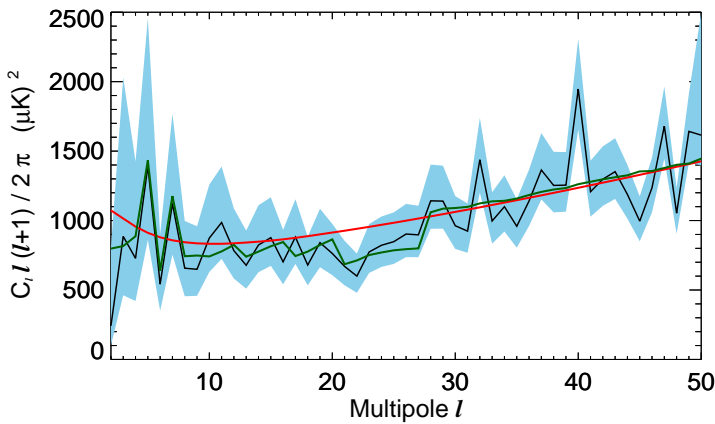}
     \caption{Angular power spectrum:  estimated $C_\ell$ from
       \textit{Planck} data (black) \citep{powspec2015};  mean $C_\ell$ of 1000
       non-Gaussian simulations (green); and  $C_\ell$ of the theoretical
       best fit $\mathrm{\Lambda CDM}$ model of 
       \citet{param2015} (red). The shaded area presents $2\sigma$ error bars from
       \citet{powspec2015}.}
     \label{fig:cl}
\end{figure}

In Fig.~\ref{fig:cl} we show the mean power spectrum of these
simulated maps compared to the \textit{Planck} best fit theoretical
$\mathrm{\Lambda CDM}$ model  \citep{param2015} and the
estimated \textit{Planck} power spectrum. We clearly see, as expected
from the construction of the filters, that the new model is in better
agreement with the data for low multipoles. In particular the low
power spectrum (anomaly A4) and the even-odd asymmetry (anomaly A6)
are evident for some multipoles. For $\ell>50$ the mean of the
simulated spectra and the best fit model are almost identical and are
not shown.

\begin{figure*}[htbp]
\centering
   \hspace*{-0.8cm}\includegraphics[width=0.4\linewidth,angle=270]{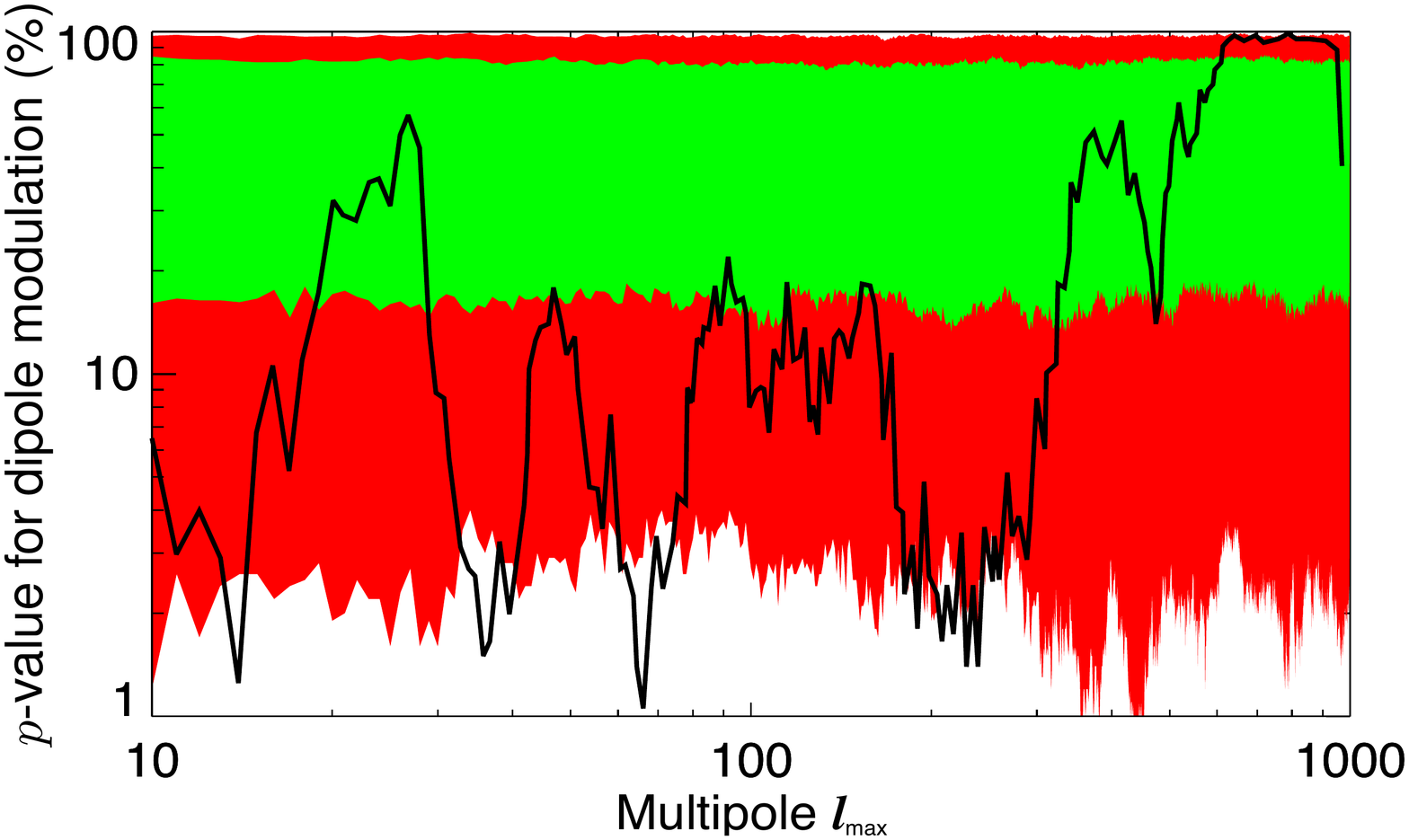}\hspace*{-1cm}\includegraphics[width=0.4\linewidth,angle=270]{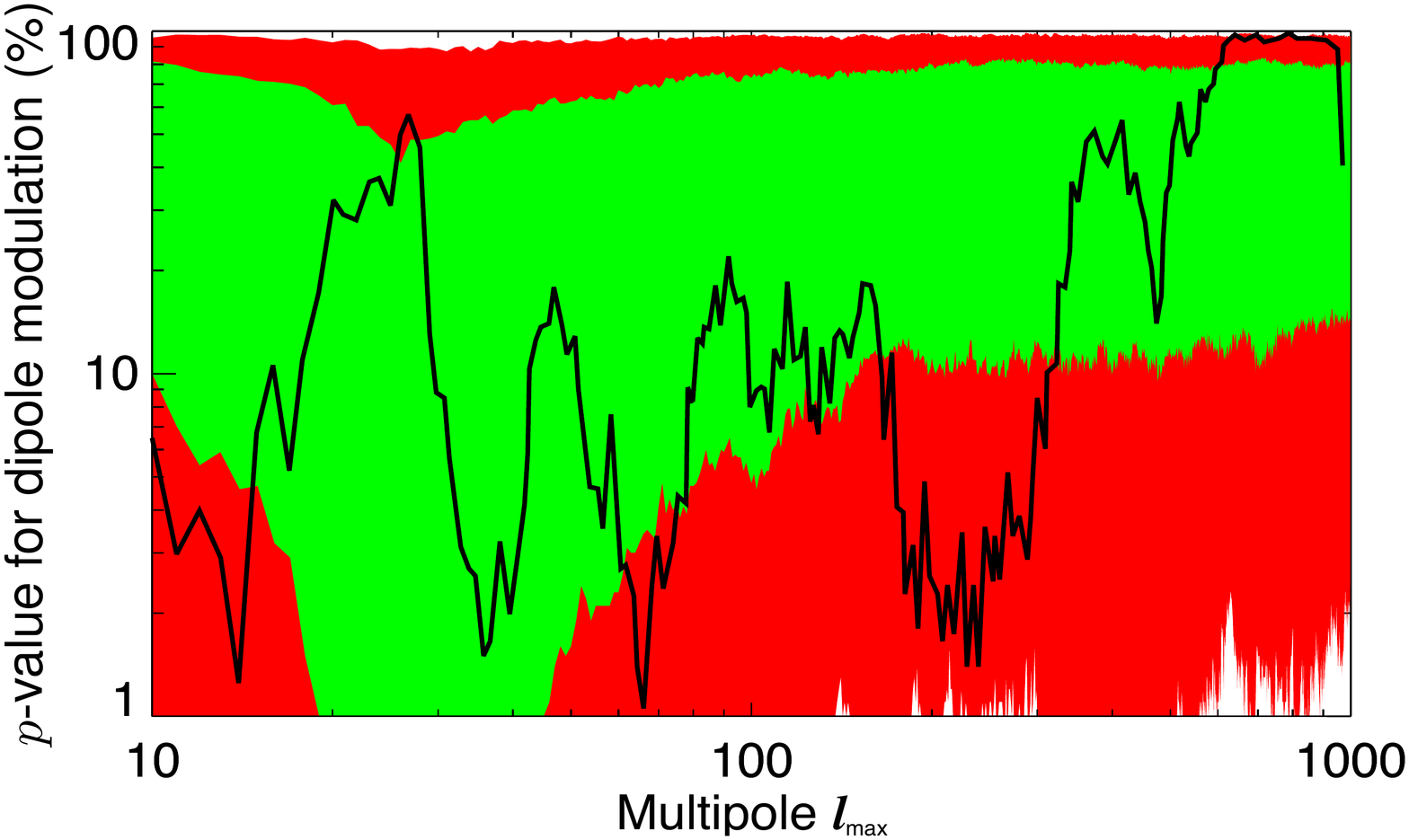}
   \caption{$p$-values for dipolar modulation, to be compared with
     figure~30 in \citet{iands2015}. Green and red bands show the
     $1\sigma$ and $2\sigma$ spread of $p$-values measured in 1000
     Gaussian simulations (left plot) and 1000 non-Gaussian model
     simulations (right plot). In both plots, the black line shows the
     $p$-values for \textit{Planck} data taken from figure~30 in
     \citet{iands2015}. These $p$-values show the percentage of
     Gaussian simulations having larger dipolar modulation up to the
     given multipole than the \textit{Planck} data (calibrated with 1000
     Gaussian simulations).\label{fig:dipmod}}
\end{figure*}

In Fig.~\ref{fig:dipmod}, we show the probability of the dipole
modulation amplitude in a given map as a function of multipole. 
In order to make the analysis of dipolar modulation
computationally feasible, the simulations for this anomaly were
analysed without using a mask. However, since the first set of 1000 simulations
used to calibrate the $p$-values is treated in the same way, we do
not expect the results to be biased. We note that the calculations were performed
following the description in \citet{iands2015}, but without the mask.
For all other analyses in this paper, we performed the analysis for the data
and simulations with the same pipeline and, as shown in the figure, 
obtain results consistent with previous papers.

Figure ~\ref{fig:dipmod} corresponds to figure~30 in \citet{iands2015}. The green and
red areas show the $68\%$ and $95\%$ intervals from Gaussian
simulations (left panel) and our toy model simulations (right
panel). The black line corresponds to the \textit{Planck} result from
\citet{iands2015}.  The left panel shows that the $p$-values for the
data are outside the $68\%$ interval for almost all multipoles
$\ell<200$ compared to Gaussian simulations. In addition there are
several dips outside the $95\%$ interval.
Conversely, the \textit{Planck} data points seem consistent with our toy model,
as shown in the right panel.  The clear dip of the $68\%$ green range
for $\ell<100$ indicates that a strong dipolar modulation is expected
on large angular scales in this model.

\begin{figure*}[htbp]
\centering
   \hspace*{-0.8cm}\includegraphics[width=0.4\linewidth,angle=270]{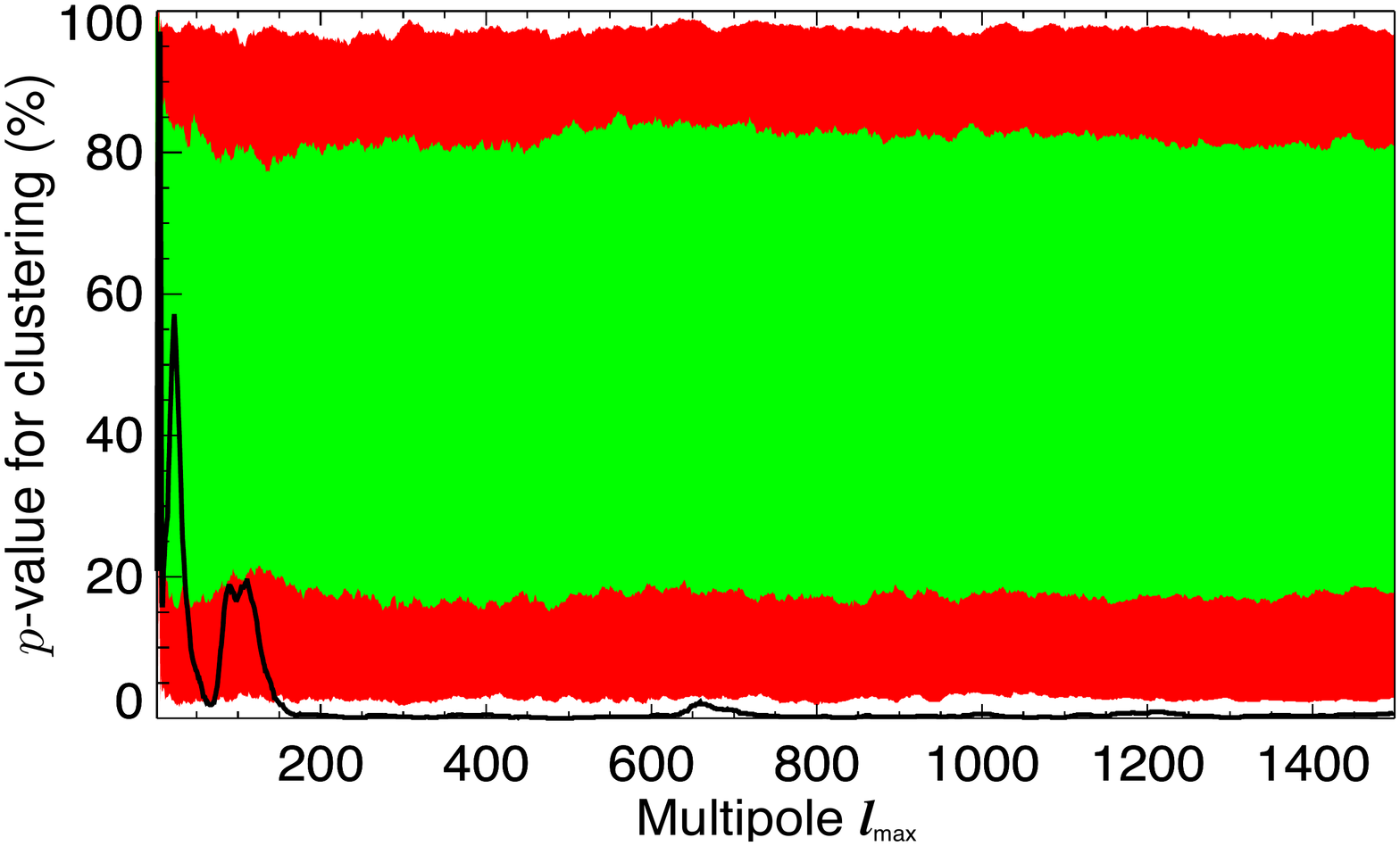}\hspace*{-1cm}\includegraphics[width=0.4\linewidth,angle=270]{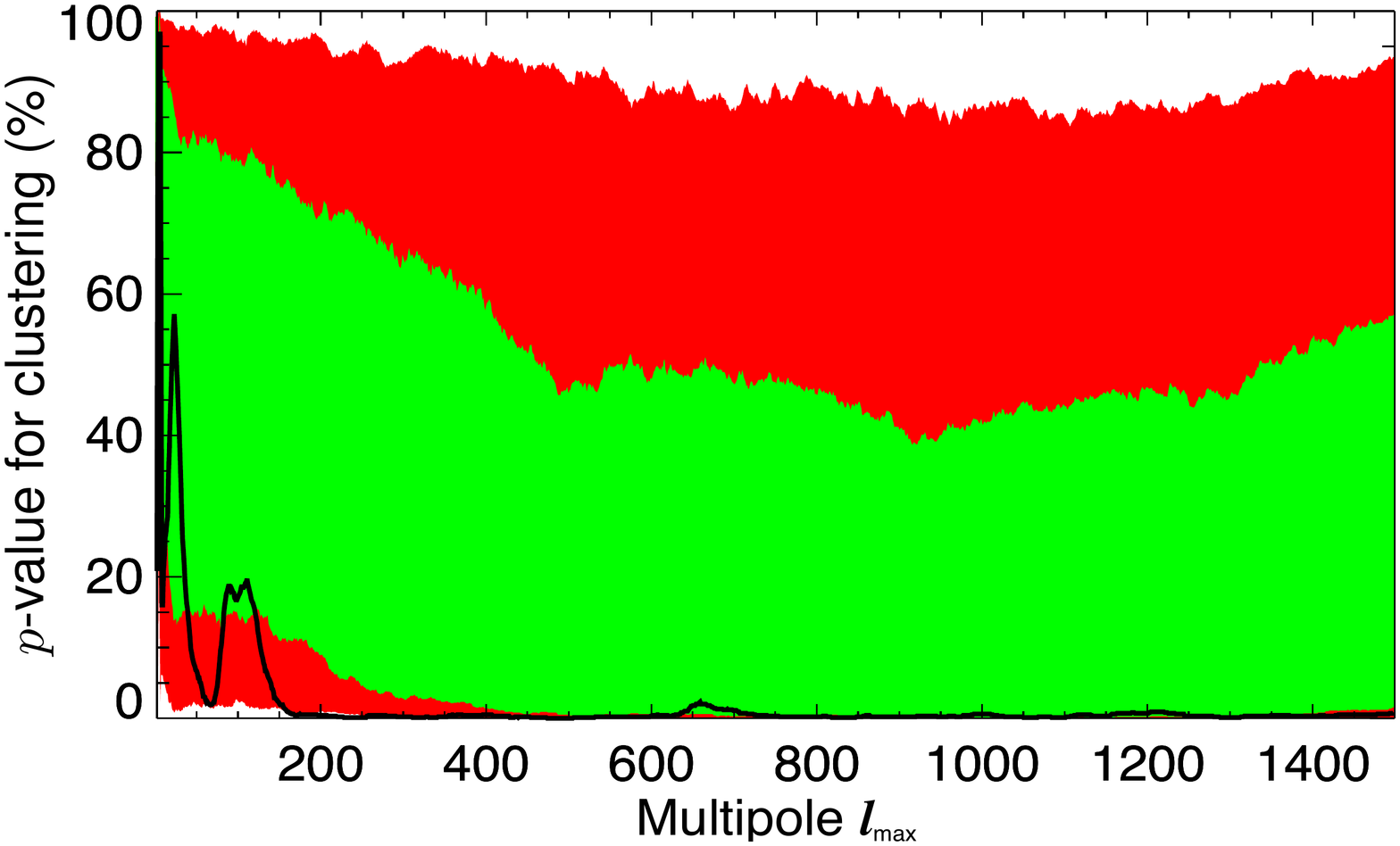}
   \caption{$p$-values for alignment of the spatial power distribution
     dipoles up to a given multipole. This is the equivalent plot to
     figure~36 in \citet{iands2015}. The green and red bands indicate the
     $1\sigma$ and $2\sigma$ spread of $p$-values measured in 1000
     Gaussian simulations (left panel) and 1000 non-Gaussian model
     simulations (right panel). The black line corresponds to the
     $p$-values for \textit{Planck} data taken from figure~36 in
     \citet{iands2015}. These $p$-values show the percentage of
     Gaussian simulations with a larger alignment of the spatial power
     distribution up to the given multipole (calibrated with 1000
     Gaussian simulations). \label{fig:alignment}}
\end{figure*}

In Fig.~\ref{fig:alignment} we show the probability of alignment of
the power distribution dipoles up to a certain multipole (compare  Fig.~36 in \citealt{iands2015}).  The green and red areas
show the $68\%$ and $95\%$ intervals from Gaussian simulations (left
panel) and our toy model simulations (right panel). The black line
shows the results determined from the \textit{Planck} data and has been taken
from figure~36 in \citet{iands2015}. In the left panel, it can be seen
that the data, as compared to Gaussian simulations, always lie outside
the $95\%$ interval for $\ell>200$. The right panel demonstrates that
the behaviour of the data is consistent with the non-Gaussian
toy model.

\begin{figure*}[htbp]
\centering
   \hspace*{-0.8cm}\includegraphics[width=0.55\linewidth]{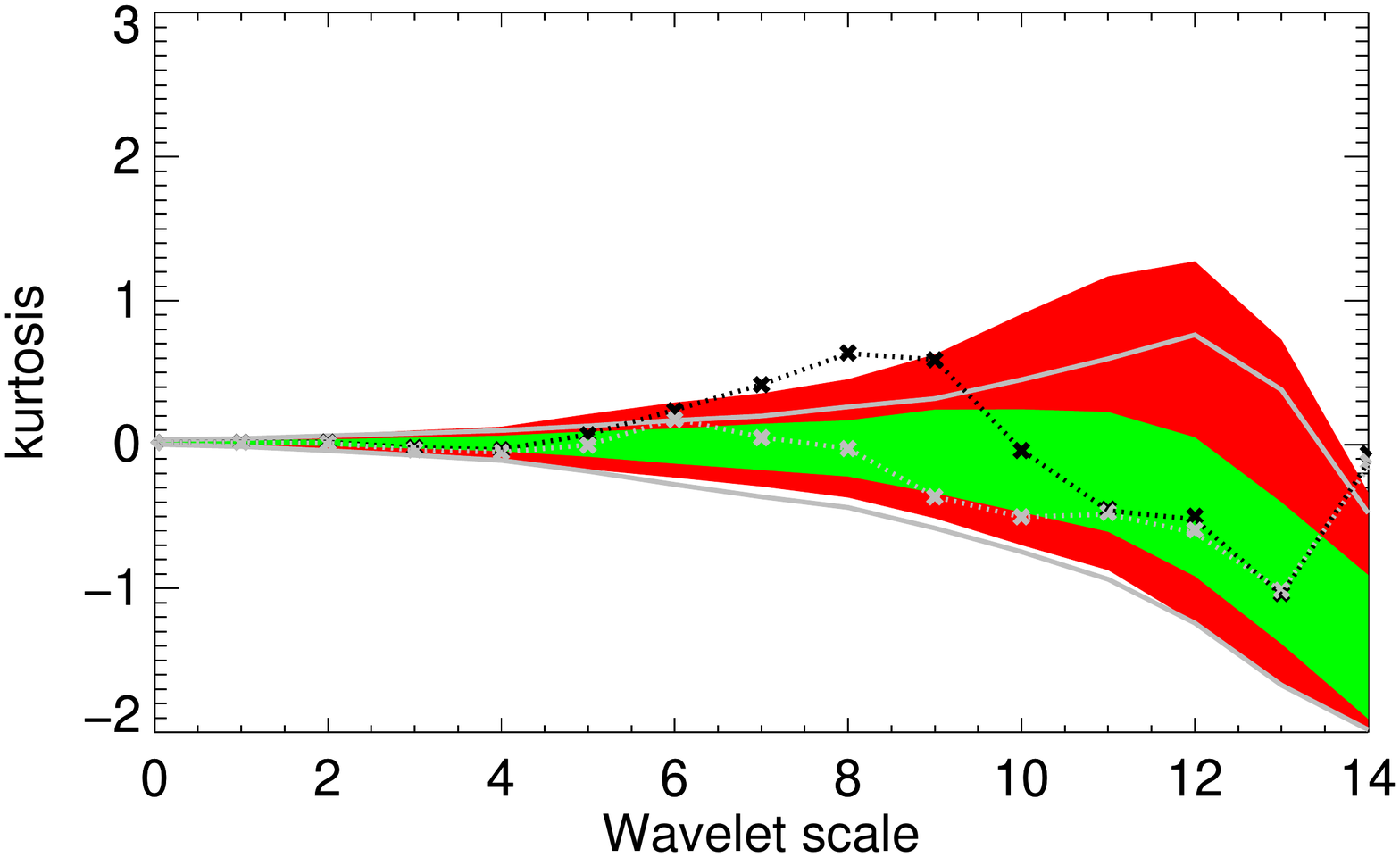}\hspace*{-1cm}\includegraphics[width=0.55\linewidth]{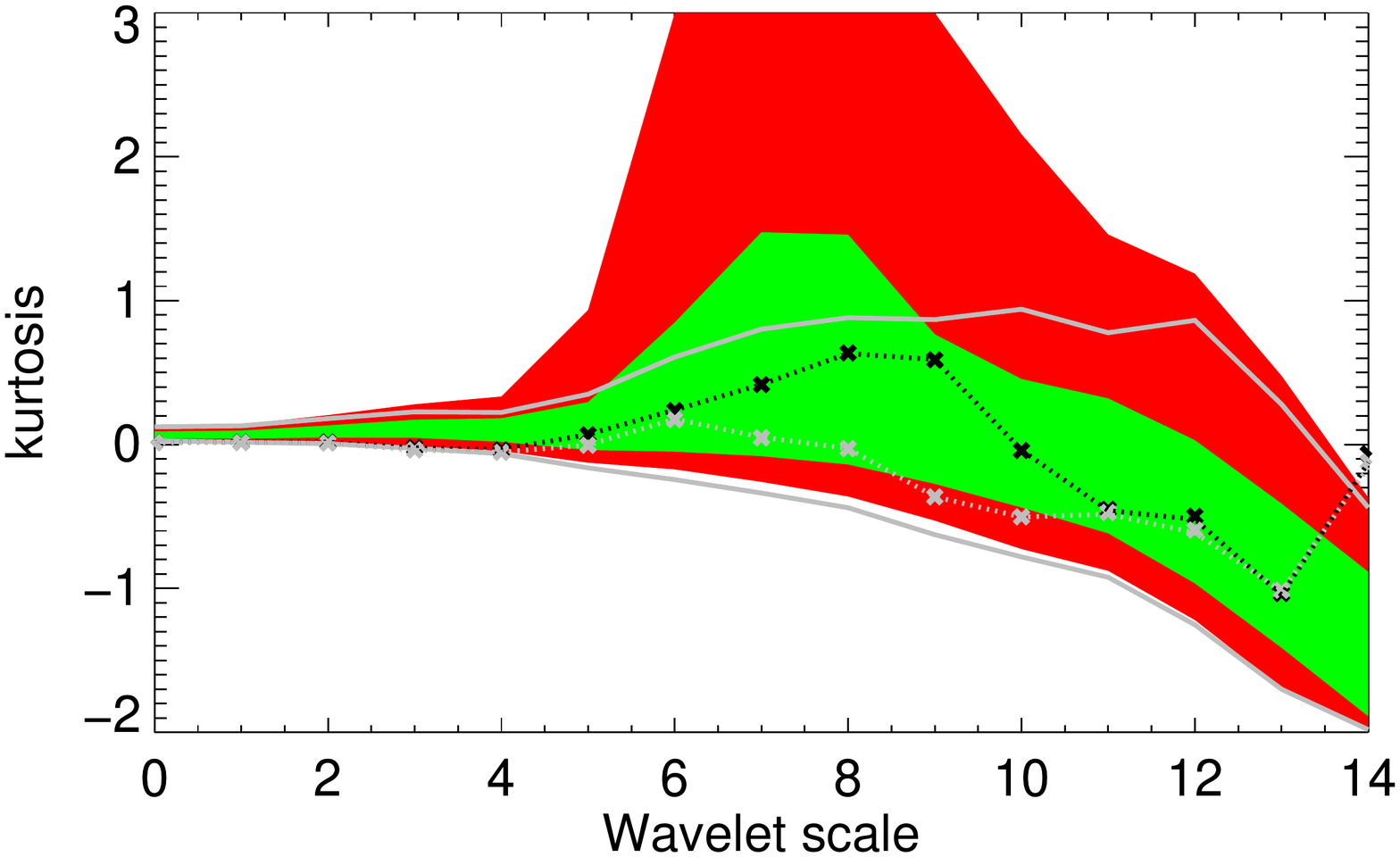}
   \caption{Kurtosis of wavelet coefficients for the wavelet scales
     defined in \citet{vielva}. Green and red bands show the $1\sigma$
     and $2\sigma$ spread of kurtosis values in 1000 Gaussian
     simulations (left panel) and 1000 non-Gaussian model simulations
     (right panel). The black crosses indicate the kurtosis values
     computed from the \textit{Planck} data; the grey crosses show the values
     derived after masking the brightest spot for the given wavelet
     scale. The grey lines show the $2\sigma$ limits after masking the
     brightest spot for the given scale. The masking is performed with
     a disc of radius 5 degrees.  \label{fig:kurtosis}}
\end{figure*}

Figure~\ref{fig:kurtosis} shows the kurtosis for wavelet coefficients
using spherical Mexican Hat wavelets and the same wavelet scales as in
\citet{vielva}. The left panel shows the kurtosis compared to Gaussian
simulations (green and red shaded bands). For scales 7-9 the \textit{Planck}
data show excess kurtosis outside the $95\%$ confidence region. In the
toy model (right panel), we see a clearly enhanced probability for an
excess kurtosis at exactly these scales. The data points for scales
7-9 are now within the $68\%$ confidence region. The scale-dependent
kurtosis of wavelet coefficients also put limits on a possible
scale-dependence of a $g_\mathrm{NL}$ non-Gaussianity. The plot shows
that our model predicts a level of kurtosis consistent with current
observational constraints.

\citet{vielva} have shown that the excess observed kurtosis disappears
after masking the highest temperature outlier in the map. This is
shown in Fig.~\ref{fig:kurtosis} where the grey crosses represent the
kurtosis values computed from the data after masking, and the grey
lines indicate the 2$\sigma$ confidence interval after masking the
simulations. For the toy models, there is a clear drop in kurtosis
after masking the brightest spot showing that the excess kurtosis in
the toy model simulations is indeed mainly associated with one strong
hot or cold spot, as for the observational data.

\begin{figure*}[htbp]
\centering
   \hspace*{-0.8cm}\includegraphics[width=0.55\linewidth]{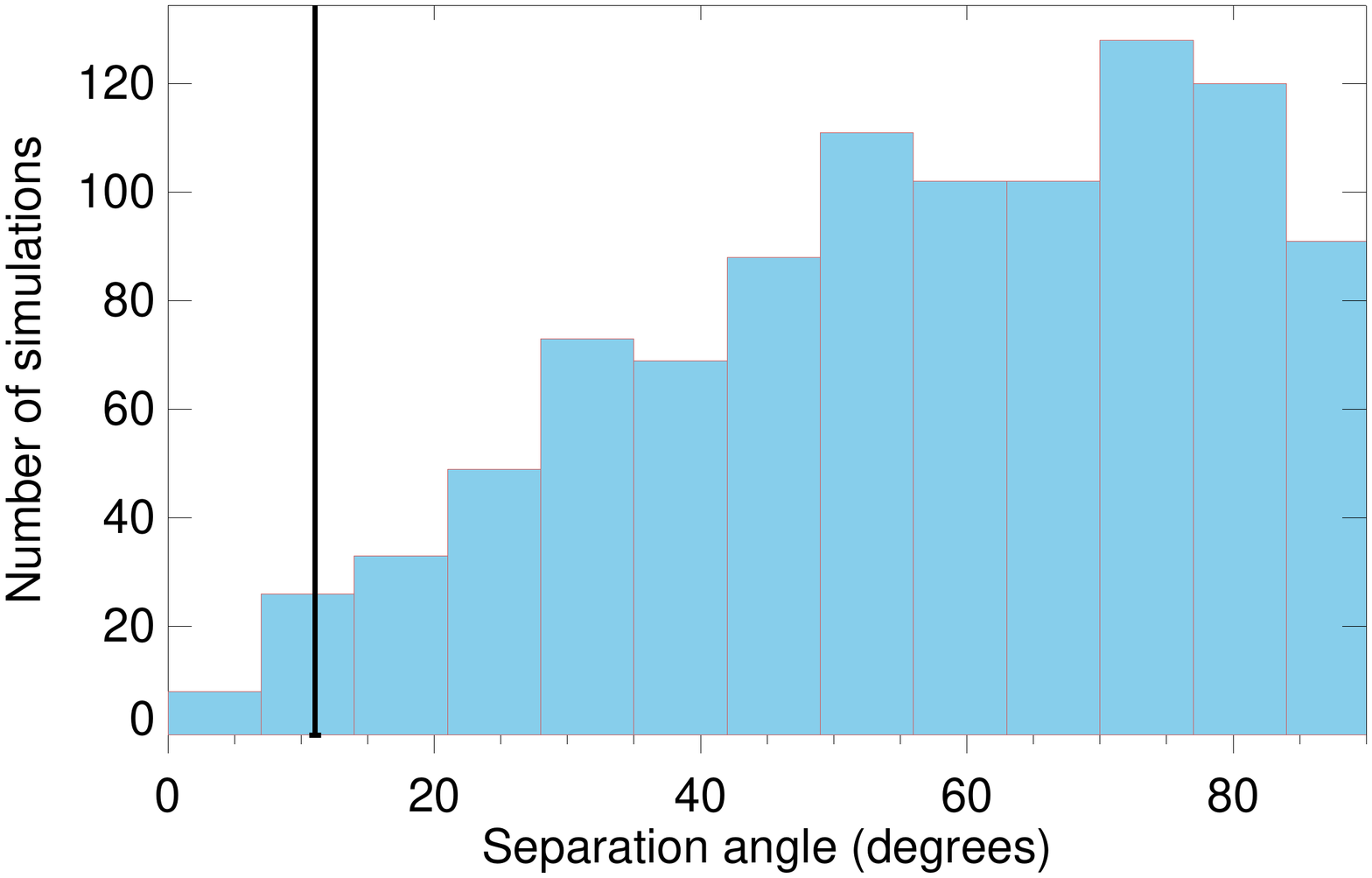}\hspace*{-1cm}\includegraphics[width=0.55\linewidth]{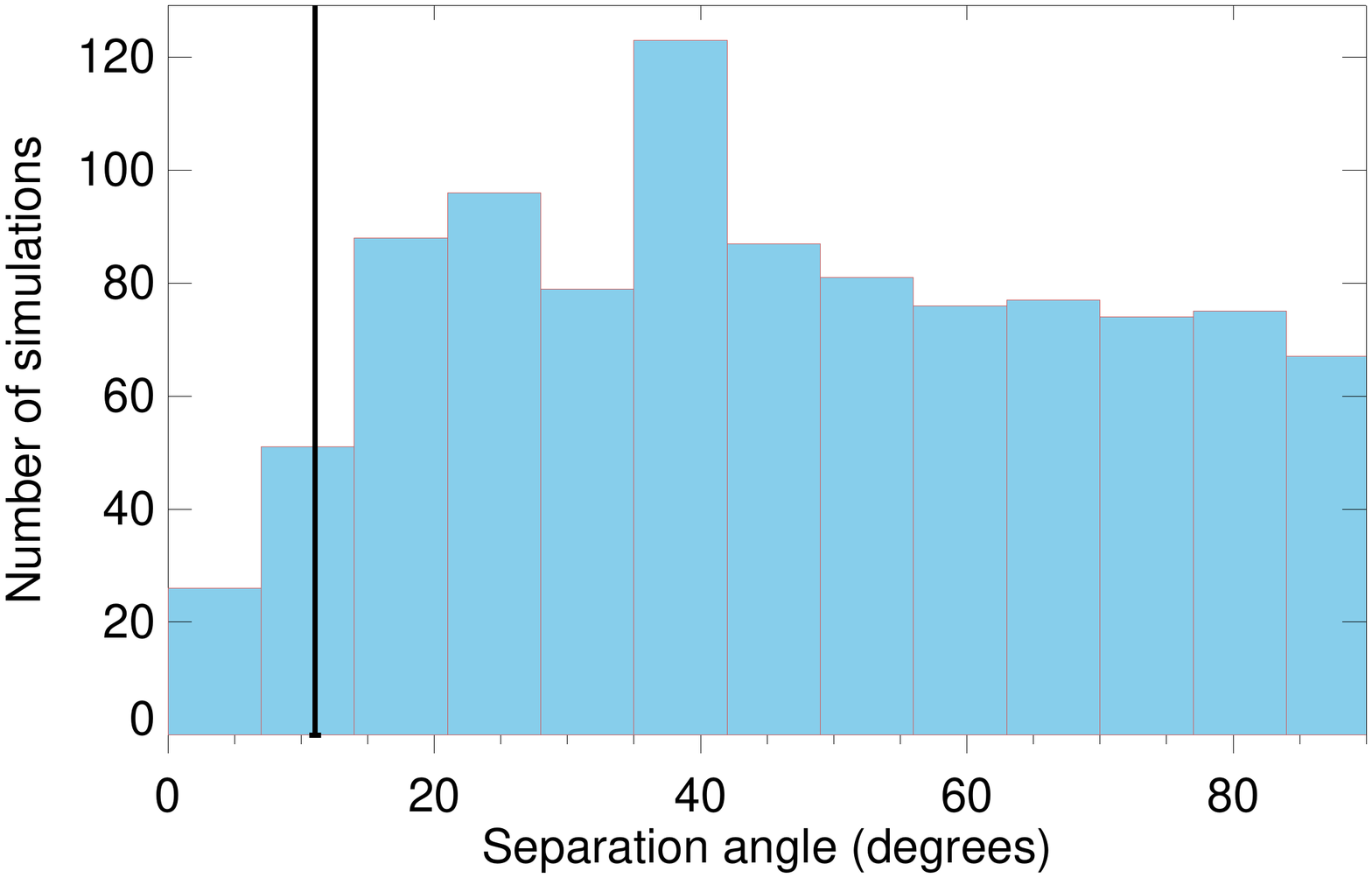}
   \caption{Angular separation between the quadrupole and octopole
     preferred directions. The left panel shows the distribution of
     this angular distance for 1000 Gaussian simulations, the right
     panel shows the corresponding distribution for 1000 non-Gaussian
     simulations. The vertical black line represents the angular
     distance observed in the \textit{Planck} data.\label{fig:quadocto}}

\end{figure*}

Figure~\ref{fig:quadocto} shows how the angular separation between the
quadrupole and octopole preferred directions are distributed in
Gaussian simulations (left panel) and toy model simulations (right
panel). The vertical black line represents this angle for the \textit{Planck}
data. The left panel indicates that the probability falls for smaller
angles. For toy model simulations, the distribution is somewhat
flatter, therefore the quadrupole-octopole alignment seen in the data
can be considered  less anomalous.

\begin{figure*}[htbp]
\centering
   \hspace*{-0.8cm}\includegraphics[width=0.55\linewidth]{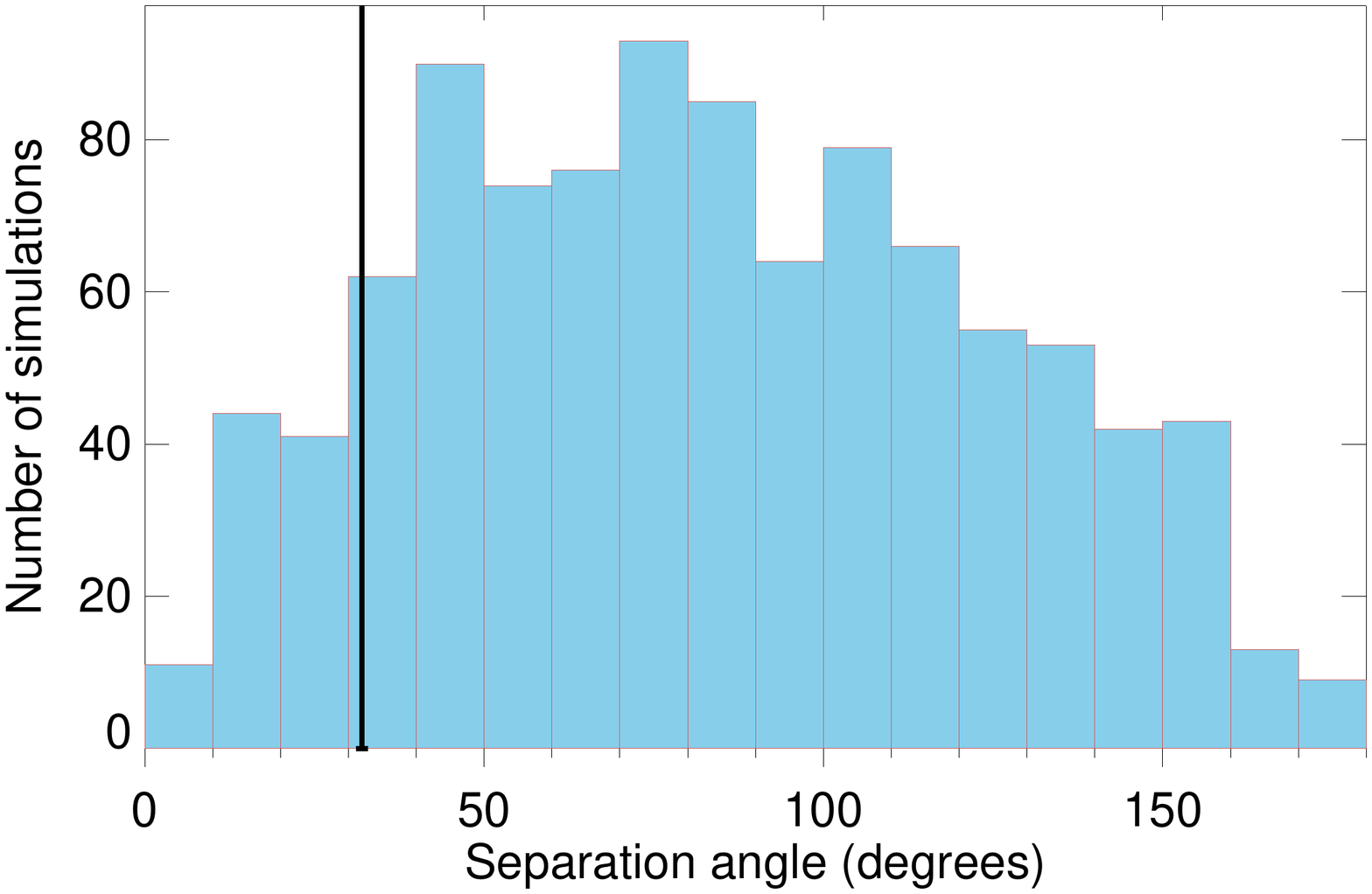}\hspace*{-1cm}\includegraphics[width=0.55\linewidth]{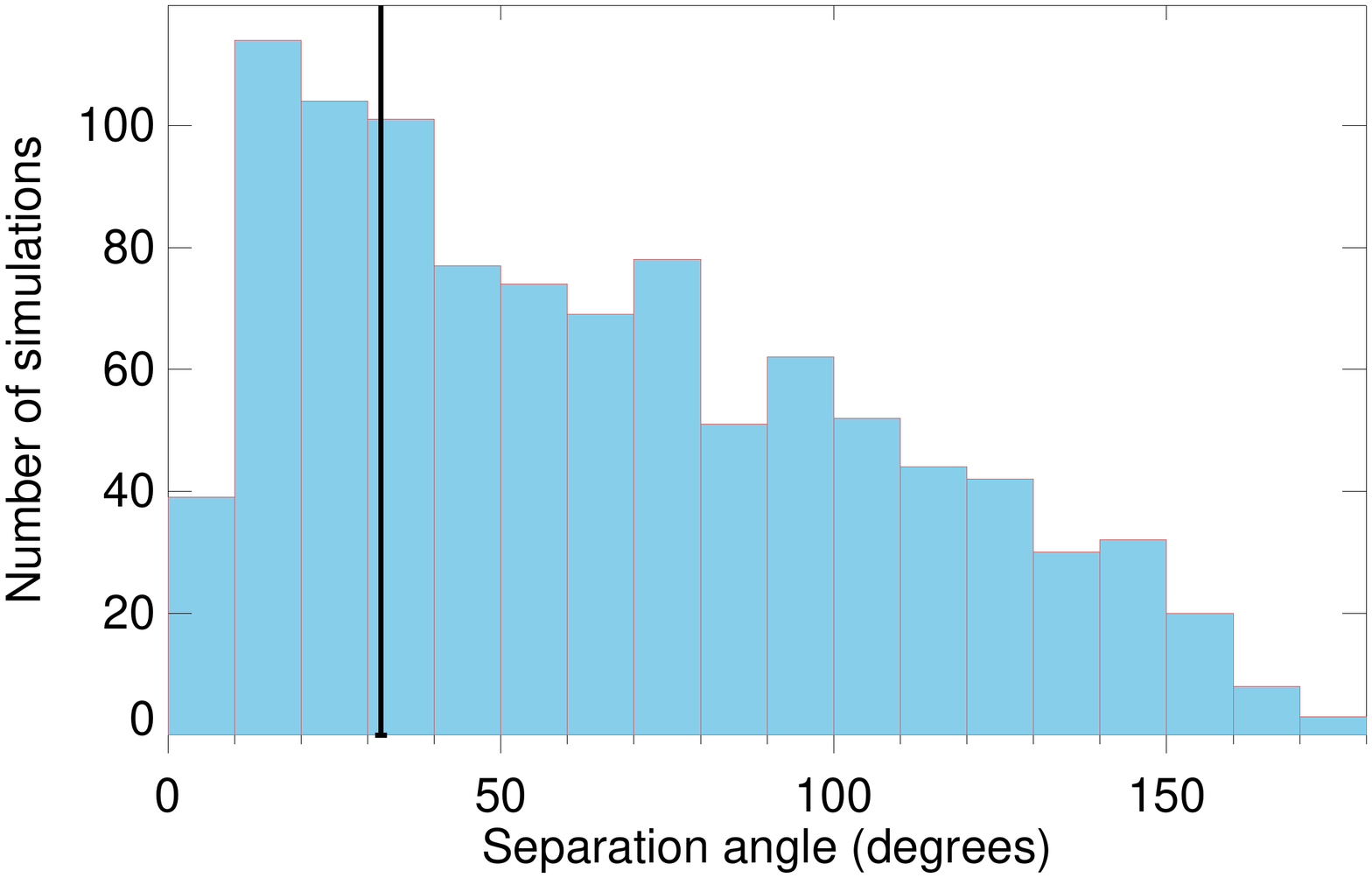}
   \caption{Angular distance between hot or cold spot and direction of
     max dipolar modulation. The distribution of the angular distance
     is shown for 1000 Gaussian simulations (left panel) and 1000
     non-Gaussian simulations (rights panel). The vertical black line
     corresponds to the angular distance for the \textit{Planck}
     data.\label{fig:spotdipang}}
\end{figure*}

Finally, the direction of dipolar modulation, the cold spot and the
directions of the alignment asymmetry all seem to be converging.  In
particular, the angular distance between the direction of dipolar
modulation and the cold spot is $32^\circ$ in the \textit{Planck} data. In
Fig.~\ref{fig:spotdipang} we show the distribution of angular
distances between the dipolar modulation and the cold/hot spot in
Gaussian simulations (left panel) or toy model simulations (right
panel). The position of the cold/hot spot is clearly strongly
correlated with the dipolar modulation direction in the toy model
simulations and in excellent agreement with the data. A similar
correlation of direction with the small-scale hemispherical asymmetry
is seen in toy model simulations with a strong alignment asymmetry.

\section{Discussion and conclusions}

In this paper we have shown that the CMB anomalies, including apparent
deviations from statistical isotropy and features in the power
spectrum can arise from non-Gaussianity. In particular, in the analyses of
simulated toy model maps using a $g_\mathrm{NL}$-like non-Gaussian
term of the form given in Eq.~\eqref{eq:model} or \eqref{eq:altmodel}, all
of the most commonly discussed anomalies are reproduced. To what
extent the different anomalies are present depends on the filters
$w_\ell$ and $g_\ell$ (which would correspond to specific
scale dependences of the primordial non-Gaussianity trispectrum
$g_\mathrm{NL}$). Even very simple forms of these filters give rise to
several anomalies in our phenomenological model, but a physical model
would be required to predict their shape with a minimum number of free
parameters.

Figure~\ref{fig:cl} demonstrates how such a toy model results in low
power on large angular scales, including the quadrupole, and parity
asymmetry for the first few multipoles.  Then, in
Figs.~\ref{fig:dipmod} to \ref{fig:quadocto}, we see evidence that
various features of the data, characterised as 2-3 $\sigma$ outliers
when compared to Gaussian simulations, are more consistent with the
expectations of these toy model simulations.

It should be noted that the filter functions selected here,
which effectively define scale dependence, are simple. Further work is
needed to assess whether these functions are realistic in the context of a
primordial underlying model or otherwise.  However, it may be that a
physical model could give rise to more complex filters and still
reproduce the anomalies if it essentially mimics  the main features
displayed by the phenomenological model focused on in this work.

We have focused on $g_{\mathrm NL}$ models here. While
$f_{\mathrm NL}$ and $\tau_{\mathrm NL}$ models may reproduce some of
the anomalies, they cannot easily reproduce all of them, whereas
$g_{\mathrm NL}$ models appear to. Anomaly A1 and possibly A2 could arise in
$\tau_{\mathrm NL}$ models \citep{shandera2018}, but these models
would not generally give a non-Gaussian hot or cold spot. For this an
enhancement of the original Gaussian fluctuations would be necessary,
which is not easily achieved in a $\tau_{\mathrm NL}$ model where the
non-Gaussian term is strongly influenced by an independent field. For
the same reason, they  also would not easily give quadrupole-octopole
alignment, where  the non-Gaussian term also needs to be strongly
correlated with the Gaussian term.

Similarly, it would be difficult for an $f_{\mathrm NL}$ model to give
large-scale power deficit and quadrupole-octopole alignment, since for
a second-order term positive fluctuations would be enhanced, while
negative fluctuations would be erased. An excess kurtosis would also
not arise from a second-order term (at lowest order in the
perturbations). As we have seen in this paper, the reason why
$g_{\mathrm NL}$ models can generate all of the anomalies originates
from the way in which the non-Gaussian term correlates to the Gaussian
term.

In our toy model (Eq.~\ref{eq:model}), we have used CMB maps that
include the effects of radiative transfer as the basis of the
non-Gaussian term.  Inspection of Eq.~\eqref{eq:gnl} indicates that, for
an inflationary model, it is the non-Gaussian third-order term in the
gravitational potential (with an overall amplitude $g_\mathrm{NL}$)
that would be transferred to the CMB anisotropies via the CMB
radiation transfer function.  In order to be more consistent with this
scenario, we  generated Gaussian maps with (1) a pure Sachs-Wolfe
spectrum and (2) a pure white noise spectrum
($C_\ell=\mathrm{const}$). We  then generated the non-Gaussian
term based on these Gaussian maps, and then applied radiative transfer
by changing the variance of the $a_{\ell m}$ coefficients of the final
maps in order to obtain a spectrum consistent with \textit{Planck} best fit. In
both cases, after modifying $g_\ell$ and $w_\ell$ accordingly, all
anomalies are again reproduced.

It should be noted  that we set the filter $w_\ell$ to zero for $\ell>27$ in order to
minimise the impact on the power spectrum for larger multipoles. The
\textit{Planck} data indicates that the power spectrum is low up to
$\ell=27$. Nevertheless, we tested the effect of (i) setting $w_\ell$
to zero for $\ell>15$ and (ii) setting $w_\ell$ to zero for $\ell>50$
with a non-zero filter extending to $\ell=50$. In both cases the
spectrum was modified to be low in the range where $w_\ell$ is
non-zero, thereby reducing the consistency with the \textit{Planck} power
spectrum.  Furthermore, the scales of the cold or hot spot and
the dipolar modulation were both altered, in general resulting in worse
agreement with the data.  We interpret this as an indication of
correlation between the angular scales where the power spectrum is low
and the angular scales of dipolar modulation (A1) and the cold spot
(A3). A similar correlation is seen in that the filters that cause
quadrupole-octopole alignment (A5) also give a low quadrupole.  We note,
however, that there is considerable freedom in the way that the
filters can be adjusted; therefore, we restricted our investigations to
simple extensions of the toy model used here.

In Figs.~\ref{fig:dipmod} to \ref{fig:quadocto}, we  show how the
anomalies are easily reproduced by the toy model simulations. However,
in general, not all anomalies will be present in a given simulation;
there is considerable variation in terms of which anomalies are seen
from realisation to realisation, and some simulations do not show
clear signs of any anomalous behaviour.  Furthermore, as shown in
Fig.~\ref{fig:maps}, the non-Gaussianity, and thereby the anomalies,
may only be visible in localised parts of the sky that are either
partially or fully rejected from further analysis by the application of
a suitable Galactic mask.  These issues make it difficult to predict
what we should expect for polarisation data in our toy model.  If we
assume that the non-Gaussian polarisation term can be obtained through
a similar mechanism, then---given that the signal is only partially
correlated with the temperature realisation---we would not necessarily
expect the same anomalies to appear, either with the same amplitude or
a similar direction.  Indeed, without a theoretical model, we cannot
make clear predictions for what to expect in polarisation.

Finally, we reiterate that the scope of this paper was to guide
theoretical research by proposing a general form for a non-Gaussian
term that might be the origin of all of the observed CMB anomalies.  The next
step must then be to determine whether an inflationary model exists
that can reproduce the main features of the phenomenological model
proposed in this paper.  An actual physical model could
take a slightly different form with different filters and still
reproduce the anomalies. Then, only when a physically motivated model
is found can a complete comparison to data be undertaken, and
predictions made for other anomalies and possible features in the CMB
polarisation signal.  This should help to alleviate the `multiplicity
of tests' arguments \citep{DPH2008,Contreras2017} made against claims
of anomalies in the data.

\begin{acknowledgements}
  The results in this paper are based on observations obtained with \textit{Planck}
  (http://www.esa.int/Planck), an ESA science mission with instruments
  and contributions directly funded by ESA Member States, NASA, and
  Canada. The simulations were performed on resources provided by
  UNINETT Sigma2 - the National Infrastructure for High Performance
  Computing and Data Storage in Norway. NB and ML acknowledge partial
  financial support by ASI Grant No. 2016-24-H.0. Maps and results have been derived
  using the Healpix software package developed by \cite{healpix}.

\end{acknowledgements}

\end{document}